\documentclass[11pt,twoside]{article}

\usepackage{asp2006}
\usepackage{epsf}
\usepackage{psfig}
\usepackage{lscape}

\begin{document}
\title{GPS and CSS radio sources and space-VLBI}   
\author{I. Snellen}   
\affil{Leiden Observatory, Leiden University, Postbus 9513, 2300 RA Leiden, The Netherlands}    

\begin{abstract} 
A short overview is given of the status 
of research on young extragalactic radio sources. We concentrate on 
Very Long Baseline Interferometric (VLBI), and space-VLBI results obtained
with the VLBI Space Observatory Programme (VSOP). In 2012, VSOP-2 will be
launched, which will allow VLBI observations at an unprecedented angular
resolution. One particular question VSOP-2 could answer is whether some of 
the High Frequency Peakers (HFP) are indeed the youngest objects in the family of 
GPS and CSS sources. VSOP-2 observations can reveal their angular morphology
and determine whether any are {\sl Ultra-compact Symmetric Objects}.
\end{abstract}


\section{Introduction: nomenclature for young radio sources}

Historically, young radio sources are named according to their observational 
characteristics. This has evolved into
the use of several 'classes' of object, which unfortunately does not make this 
area of research more transparent. Furthermore, the introduction of new names or 
classes of object (as will be done in this paper) is not 
necessarily accompanied by a deeper physical understanding of the subject.
Nevertheless, it is probably a good idea to briefly summarize the observational
properties of the different classes and their possible connections.

Gigahertz Peaked Spectrum (GPS) sources are compact (typically 10-100 m.a.s.)
radio sources characterized by their convex spectrum peaking around 
1 GHz in frequency. Their spectral turnover is thought to be due to 
synchrotron self absorption (e.g. Fanti et al. 1990; Snellen et al. 2000), 
although free-free absorption may also play a role in some objects (e.g. Sawada-Satoh et al., 2002). 
Flux density variations in GPS sources are generally found to be small, 
in particular 
in those sources optically identified with galaxies. GPS quasars are
found at much higher redshifts (z$>$2), often show more significant 
variability, having core-jet morphologies. They may not be related to 
the GPS galaxies and just be a sub-class of flat spectrum quasars.
A GPS galaxy typically exhibits a two-component structure with sometimes a weak 
flat spectrum component in between, interpreted as the core.
If such an object is discovered (or observed) with VLBI, it is
referred to as a Compact Symmetric Object (CSO). In particular if 
the overall radio spectrum has not yet been determined. Ignoring one or 
two galaxies (and the GPS quasars), GPS and CSO are the same beasts.
All famous and well studied GPS galaxies are CSO, and vice verse. A powerful
radio source of a few tens of milliarcseconds in size will always be synchrotron
self absorbed, and when it is observed at a large viewing angle, the lack of 
significant Doppler boosting will allow both sides
of the radio source to be seen (CSO), which will also result in low flux variability 
and a stable GHz-peaked spectrum (GPS).

Compact Steep Spectrum (CSS) sources are larger than GPS sources, up to 1 or 2 arcseconds
in size, and are characterized by steep spectra (in contrast to other
types of compact radio sources). Their spectra typically turn over at about 100 MHz. This is consistent  
with synchrotron self-absorption, with the spectral turnover of a radio source depending on angular size as $\theta^{-4/5}$. The morphologies of CSS sources seem more complex than
those of GPS sources, often showing bends and twists, or bright knots in their jets. 
This may be biased by the fact that there are more
resolution elements available over the size of the source. Furthermore, those sources 
with strong interaction with their galactic environments will be intrinsically brighter
and therefore more prominent in the bright end of the CSS population.
Those CSS sources for which both sides of the radio source jets/lobes are observed are
sometimes called Medium Symmetric Objects (MSO).
 
Recently, a new 'class' has been defined in {\sl young radio source land}, the High Frequency Peakers (HFP; Dallacasa et al. 2000). While CSS sources are peaking at low frequencies and are 
significantly larger than GPS sources, HFP have a spectral turnover frequency at
$>$5 GHz, and are significantly more compact than GPS. As will be discussed below, 
there is compelling evidence that GPS/CSO and CSS/MSO sources fit into an evolution 
scheme, where GPS sources are 10$^{2-3}$ years old and grow into their CSS phase
at an age of 10$^{4-5}$ years. It is the question whether HFP also fit into 
this scheme, at the beginning, as the youngest of the whole family. It is clear that 
the HFP class is heavily contaminated by Blazar/BL-Lac type objects (see below), but possibly a small fraction of the HFP class are indeed very young. As will be argued in this paper, space-VLBI observations will have the required angular resolution to 
reveal their morphologies. This will show whether a sub-sample of these sources 
exhibit double-lobed morphologies, which we here dub (to further confuse the nomenclature 
in this field) Ultra-compact Symmetric Objects (USO).

\section{Determination of radio source ages}

Over the last two decades, compelling evidence has been accumulated that GPS/CSS
sources are indeed young objects. This was first proposed by Shklovsky in 1965,
although in that same period it was also investigated whether the 
radio emission of the archetypical GPS source, B1934-63, was
actually a signal from a possible extraterrestrial civilization (Kellermann 1966).
The first important argument is that the spectral turnovers of these radio sources
must mean that there is no (or hardly any) radio emission at large scales. Any large scale
emission present would not be synchrotron self-absorbed and dominate the flux
at low frequencies. This in contrast to most other compact radio sources
which have {\sl flat} spectra, implying the presence of components with a large range 
of physical scales in these objects. Furthermore, GPS sources were found to exhibit 
low or no flux density variability. This indicates that relativistic Doppler boosting 
(as found in Blazars and BL-Lac objects) seems not to be important, implying
that these objects do not have a small angular size because they just happen to 
be observed at a small viewing angle. This was 
consistent with early VLBI observations which often showed a double structure
interpreted as two mini-lobes (Philips \& Mutel 1982). The youth argument was 
further strengthened by the detection of cores or core/jet structures in 
the centers of these objects, identifying them as CSO (Wilkinson et al 1994; 
Conway et al. 1994). However, the most compelling evidence, putting the 
youth interpretation beyond reasonable doubt, comes from VLBI measurements of 
the angular separation velocity of the lobes in several of the most extensively 
studied GPS/CSO (Owsianik \& Conway 1998; Owsianik, Conway, \& Polatidis 1998;
Tschager et al. 2000; Polatidis \& Conway 2003). They reveal expansion 
velocities of 10-20\% of the speed of light, implying kinetic source ages for 
GPS sources of a few hundred to a few thousand years. The ages of the larger CSS
sources are estimated using the breaks in their synchrotron spectra, and found to 
be in the order of 10$^{4-5}$ years (Murgia et al. 1999). 
Although these are indirect age measurements,  
they are perfectly in line with the larger size of the CSS compared to the GPS
assuming similar expansion speeds. Note that the CSS sources are too large and too
old to measure their ages directly by means of their expansion. 

\section{Young radio sources in a broader astrophysical context}

Now we know that GPS/CSS sources are young, what can we learn about them that 
is interesting in a broader astrophysical sense? Three main areas of interest
are identified:
\begin{itemize}
\item \underline{1) Long-term evolution of AGN activity.}
First of all, GPS/CSS sources, combined with samples of large size classical FR\,I and II radio galaxies, will
shed light on the growth and long-term evolution of AGN activity and their 
associated radio sources. Questions that can be raised are, what is the 
luminosity/energy evolution of a radio outburst? How long do these outbursts last?
Do all radio outburst at some point evolve in to classical radio sources, or are
some outbursts short-lived? And, how often do galaxies undergo cycles of AGN activity?
Note that X-ray or optical AGN-related emission
provides information on the current state of the AGN activity, while the radio lobes
can be tens of millions years of age, storing information from a significant part
of the AGN life-cycle (e.g. Vink et al. 2006).
\item \underline{2) Trigger and onset of AGN activity.} The ages of GPS sources
are only 10$^{2-3}$ years, a time scale so short that it is insignificant compared to
galactic time scales. This means that when we study the host galaxies and environments,
we study them at the moment they become active. This provides a unique opportunity to
probe the trigger and onset of the AGN activity, as proposed to be caused by gas-rich galaxy mergers.  
\item \underline{3) AGN feedback mechanisms.} A hot topic in current astrophysics
is the influence of AGN activity on the evolution of the host galaxy as a whole.
It has been suggested that AGN outflows quench star formation and regulate the 
stellar populations in massive elliptical galaxies (eg. Silk \& Rees 1998; Bundy et al. 2008).
The small physical extent (and young age)
of a GPS radio source means that the influence of the central AGN activity
on its galactic environment is confined to the inner (few) hundred parsec. We therefore
have a clear picture of the galactic environment free of AGN interference. By comparing
this to the host galaxies with older central AGN, the influence of the AGN on the 
host galaxy can be distilled.
\end{itemize}

It is clear that the answers to these questions can not be found using radio 
observations alone, but they need studies that make use of observations across the 
electromagnetic spectrum (e.g. de Vries et al. 2007).
In figure 1, the galactic environment of the distant
(classical) radio galaxy 3C324 is compared with that of GPS galaxy 2352$-$0604. The host
galaxy of 3C324 shows bright UV/optical emission aligned with the radio jet, possibly
caused by quasar-ionization or jet-induced star formation. Also some smaller neighbouring
galaxies seem to be visible. In the GPS galaxy 2352$-$0604 the AGN related light is not present, but clearly showing several (red) galaxy clumps in the vicinity of the GPS host galaxy.

\section{Young radio sources at the highest resolution: an outlook for VSOP-2}

The first VLBI space observations where made possible through the VLBI Space 
Observatory Programme (VSOP) with the launch of the Japanese HALCA satellite in 
1997. It operated at 5 and 1.6 GHz frequency with a factor three
improvement in angular resolution compared to ground-based global VLBI.
Since this was an engineering mission, and the sensitivity just enough to 
observe those sources with the highest surface brightnesses , only a few GPS galaxies were observed successfully. However, the VSOP observations of B2021+614 did led to the dynamical age determination of this GPS galaxies
(Tschager et al. 2000). 

VSOP-2 will be the next VLBI space mission. The satellite, Astro-G, will host a 
9 m. diameter VLBI antenna operational at 8, 22, and 43 GHz. It will be launched in
a 30,000 km altitude orbit by the Japanese Space Agency JAXA in 2012.
The great improvement compared to VSOP will be in the significant increase in sensitivity
by about an order of magnitude, and the increase in angular resolution by up to a 
factor of 5 due to the higher observing frequencies (Hirabayashi et al. 2004). 
VSOP could only observe a few
of the most compact GPS galaxies, which will be the same for VSOP-2. Although the 
sensitivity is increased dramatically, the higher observing frequency will mean that 
the GPS sources are significantly fainter. This in contrast to the general population
of compact radio sources which have flat radio spectra.
One sub-class of possibly young radio source, for which the high observing frequencies is actually
a benefit, are the High Frequency Peakers. VSOP-2 could have a major contribution
to to understanding of these objects. 

As mentioned before, the class of HFP sources is significantly contaminated by
blazar/BL-Lac type objects. In the case of blazars we observe the radio jets at 
a very small viewing angle, meaning we look right down the jet. In this 
case the jet-components
(probably shocks) have high gammas and strong relativistic Doppler boosting. When new
components emerge, they appear very bright and completely dominate the radio spectrum.
Since they are at that stage very compact and synchrotron self-absorbed, they make the
overall radio spectrum appear as that of a HFP. While the component expands, the 
turnover frequency drops, as does the flux density. For the most Doppler-boosted 
jets, which are viewed at the smallest angle to the line of sight, the time scales of 
the outbursts can be weeks or months. But for more modest viewing angles this can
be years or decades. In the latter case it is impossible to distinguish them from 
truly young radio sources. Young HFP would be expected to be less than one-hundred
years old, so they would exhibit a flux density evolution at a similar time scale 
as a HFP Doppler boosted component in a jet. VSOP-2 can distinguish between
an overall young radio source and a recently emerged jet component by determining
its morphology. While a jet component with a turnover frequency of 20 GHz would 
have a size of 100-200 $\mu$arcsec and be unresolved with the VSOP-2, a young radio 
source should exhibit a double-lobed morphology, typically $\sim$1 milli-arcseconds
in size, dubbed here an Ultra-compact Symmetric Object (USO). 
Hence the question is whether any of the HFP are USO?

\begin{figure}
\plotone{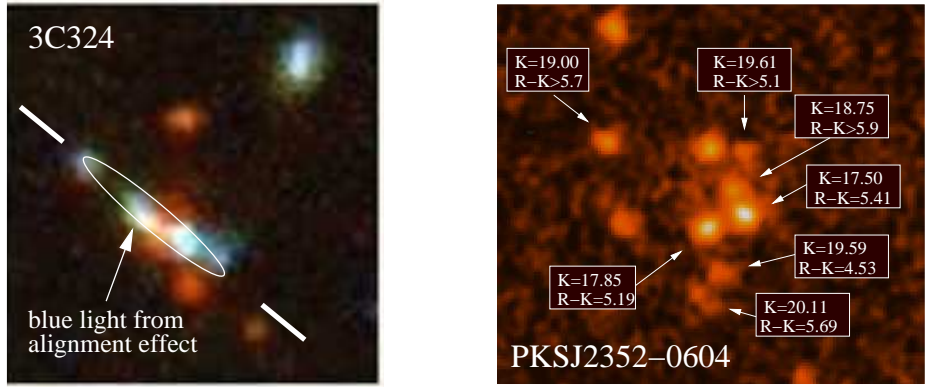}
\caption{
 The host galaxies of classical AGN, such as 3C324 (left
panel, from Subaru) are disrupted by the powerful central activity. 
The white marks
indicate the orientation of the radio jets, and the ellipse high-lights
the blue-ish aligned light
 possibly caused by ionisation, scattering, and/or jet-induced star
formation processes. The central activity in the galaxy PKSJ2339-0604 (right
panel) has just commenced. Here the effects of the AGN activity are
still confined to only the central parts of the galactic host, making
them key objects to study the environment and trigger of the AGN activity.
The host galaxy (K=17.50, R$-$K=5.41) and a further 7 surrounding
objects exhibit the defining properties of Extremely Red Objects (ERO)
with R$-$K=4$-$5.
}
\end{figure}


\acknowledgements 
I thank the organisers of the symposium for a great meeting, and for inviting 
me to give this presentation.


\end{document}